\documentclass[reprint,superscriptaddress,aps, prb,longbibliography,floatfix]{revtex4-2}
\usepackage[utf8]{inputenc}
\usepackage[T1]{fontenc}
\usepackage{amsmath}
\usepackage{amssymb}
\usepackage{graphicx}
\graphicspath{ {./Fermi/} }
\usepackage[capitalize, nameinlink]{cleveref}
\usepackage{subfigure}
\graphicspath{{figures/}}
\usepackage{adjustbox}
\usepackage{booktabs}
\usepackage{multirow}

\usepackage{color}
\usepackage{rotating}
\usepackage{ulem}
\usepackage{svg}
\usepackage{soul}
\usepackage{upgreek}
\usepackage{gensymb}

\newcommand{\supplementarysection}{
  \setcounter{figure}{0}
  \let\oldthefigure\thefigure
  \renewcommand{\thefigure}{S\oldthefigure}
  \section{Supplementary Information}
  \let\oldsection\section
  \renewcommand{\section}{
    \let\thefigure\oldthefigure
    \let\section\oldsection
    \oldsection
  }
}

\begin{document}

\title{Distinct element-specific nanoscale magnetization dynamics following ultrafast laser excitation}

\author{Emma Bernard}
\affiliation{Department of Materials Science and Engineering, University of California Davis, Davis, CA, USA}

\author{Rahul Jangid}
\affiliation{National Synchrotron Light Source II, Brookhaven National Laboratory, Upton, NY, USA}
\affiliation{Department of Materials Science and Engineering, University of California Davis, Davis, CA, USA}

\author{Nanna Zhou Hagström}
\affiliation{Department of Materials Science and Engineering, University of California Davis, Davis, CA, USA}

\author{Meera Madhavi}
\affiliation{Department of Materials Science and Engineering, University of California Davis, Davis, CA, USA}

\author{Jeffrey A. Brock}
\affiliation{Center for Memory and Recording Research, University of California San Diego, La Jolla, CA, USA}%

\author{Matteo Pancaldi}
\affiliation{Elettra Sincrotrone Trieste S.C.p.A., Area Science Park, S.S. 14 km 163.5, 34149 Trieste, Italy}%

\author{Dario De Angelis}
\affiliation{Elettra Sincrotrone Trieste S.C.p.A., Area Science Park, S.S. 14 km 163.5, 34149 Trieste, Italy}%

\author{Flavio Capotondi}
\affiliation{Elettra Sincrotrone Trieste S.C.p.A., Area Science Park, S.S. 14 km 163.5, 34149 Trieste, Italy}%

\author{Emanuele Pedersoli}
\affiliation{Elettra Sincrotrone Trieste S.C.p.A., Area Science Park, S.S. 14 km 163.5, 34149 Trieste, Italy}%

\author{Kyle Rockwell}
\affiliation{Center for Magnetism and Magnetic Nanostructures, University of Colorado Colorado Springs, Colorado Springs, CO 80918 USA}

\author{Mark W. Keller}
\affiliation{Quantum Electromagnetics Division, National Institute of Standards and Technology, Boulder, CO, USA}%

\author{Stefano Bonetti}
\affiliation{Department of Molecular Sciences and Nanosystems, Ca' Foscari University of Venice, 30172 Venice, Italy }%

\author{Eric E. Fullerton}
\affiliation{Center for Memory and Recording Research, University of California San Diego, La Jolla, CA, USA}%

\author{Ezio Iacocca}
\affiliation{Center for Magnetism and Magnetic Nanostructures, University of Colorado Colorado Springs, Colorado Springs, CO 80918 USA}

\author{Thomas J. Silva}
\affiliation{Quantum Electromagnetics Division, National Institute of Standards and Technology, Boulder, CO, USA}%

\author{Roopali Kukreja}
\affiliation{Department of Materials Science and Engineering, University of California Davis, Davis, CA, USA}%

\date{\today}

\begin{abstract}
Time-resolved ultrafast extreme ultraviolet (EUV) magnetic scattering is used to study laser-driven ultrafast magnetization dynamics of labyrinthine domains in a [Co/Ni/Pt] multilayer. Our measurements at the Co and Ni M-edges reveal distinct ultrafast distortions of the scattering pattern position and width for Ni compared to Co. Ni shows a strong modification of the scattering pattern, approximately 10 to 40 times stronger than Co. As distortions of the labyrinthine pattern in reciprocal space relate to the modification of domain textures in real space, significant differences in Co and Ni highlight a 3D distortion of the domain pattern in the far-from-equilibrium regime. 

\end{abstract}
\maketitle

Manipulating mesoscopic magnetic textures, like domains and domain walls, holds promise for energy-efficient magnetic memory and logic applications, advancing the next generation of spintronic devices \cite{Parkin,Fert2013,Caretta_dw_skyrm,Manchon}. Ultrafast laser manipulation offers a unique approach by leveraging far-from-equilibrium physics and creating novel pathways that conventional equilibrium methods cannot access \cite{Beaurepaire}. While there has been significant progress in understanding ultrafast demagnetization and spin dynamics in uniformly magnetized systems, few studies have investigated the rapid, nanoscale changes in domain patterns under ultrafast conditions \cite{ Pfau, Vodungbo,Kerber2020, Zusin, Hennes, ZhouHagstrom, Jangid, Leveille2022, Fan:22, Suturin}. Ultrafast magnetization dynamics in textured ferromagnetic systems, driven by complex interactions among charges, spins, and phonons, have challenged conventional models \cite{Battiato_38,Koopmans}. For example, recent measurements using time-resolved magnetic scattering revealed that ultrafast laser excitation can induce notable distortions in labyrinthine domain patterns and achieve domain wall velocities as high as 66 km/s \cite{Jangid}, an order of magnitude higher than predicted by Walker breakdown for standard approaches such as field or current \cite{Schryer_and_Walker, Ferré}. This observation agrees with theoretical predictions of laser-driven domain wall motion on the order of 10 km/s \cite{Balaz}. However, deeper insights into how textured systems respond to ultrafast optical stimulation are needed to harness these dynamics for practical applications and to establish theoretical models that fully describe the phenomena.

Most studies investigating the laser modification of nanoscale magnetic textures have primarily measured the dynamics of a single elemental layer in materials with perpendicular magnetic anisotropy, PMA, which facilitates the formation of labyrinthine or stripe domains. These domains are typically assumed to be separated by Bloch-type domain walls that are homogeneous through film thickness \cite{Zusin,Hubert_Schäfer}, but are understood to have deviations close to the surface or interface due to boundary conditions \cite{Hubert_Schäfer}. For multilayer PMA materials, each layer is sufficiently thin (0.5–2~nm) such that adjacent spins are coupled by exchange interactions. It has already been demonstrated that the magnetization quench of uniformly magnetized materials can be element-specific in magnetic alloys \cite{Yamamoto_2019, Vaskivskyi, Willems, Hofherr,Mathias}. However, the element-specific response of nanoscale magnetic textures across the multilayer film thickness has been missing so far. 

In this Letter, we present time-resolved small-angle magnetic resonant scattering measurements of a [Co/Ni/Pt] multilayer with PMA probed at the Co and Ni M$_{3}$ edges. Our results reveal distinct dynamics for the Co and Ni subsystems. We observe that the modification of the scattering pattern is larger for Ni than for Co. Specifically, the ultrafast change of the scattering pattern's $q$-position, related to modifications of domain wall curvature in the domain pattern \cite{Jangid}, is an order of magnitude higher in Ni than in Co. The distinct dynamics for Co and Ni reveal that the domain wall distortions across the film thickness are heterogeneous. These distortions in the Ni scattering pattern recover on an extremely fast timescale of 300-800 fs, and could be driven by the modification of exchange interactions or magnon dynamics following laser excitation. Our results provide experimental evidence of a three-dimensional (3D) deformation of the domain pattern, which needs to be considered to accurately describe mesoscale magnetization dynamics in far-from-equilibrium conditions.

\begin{figure}[t!]
    \includegraphics[width=1.0\linewidth]{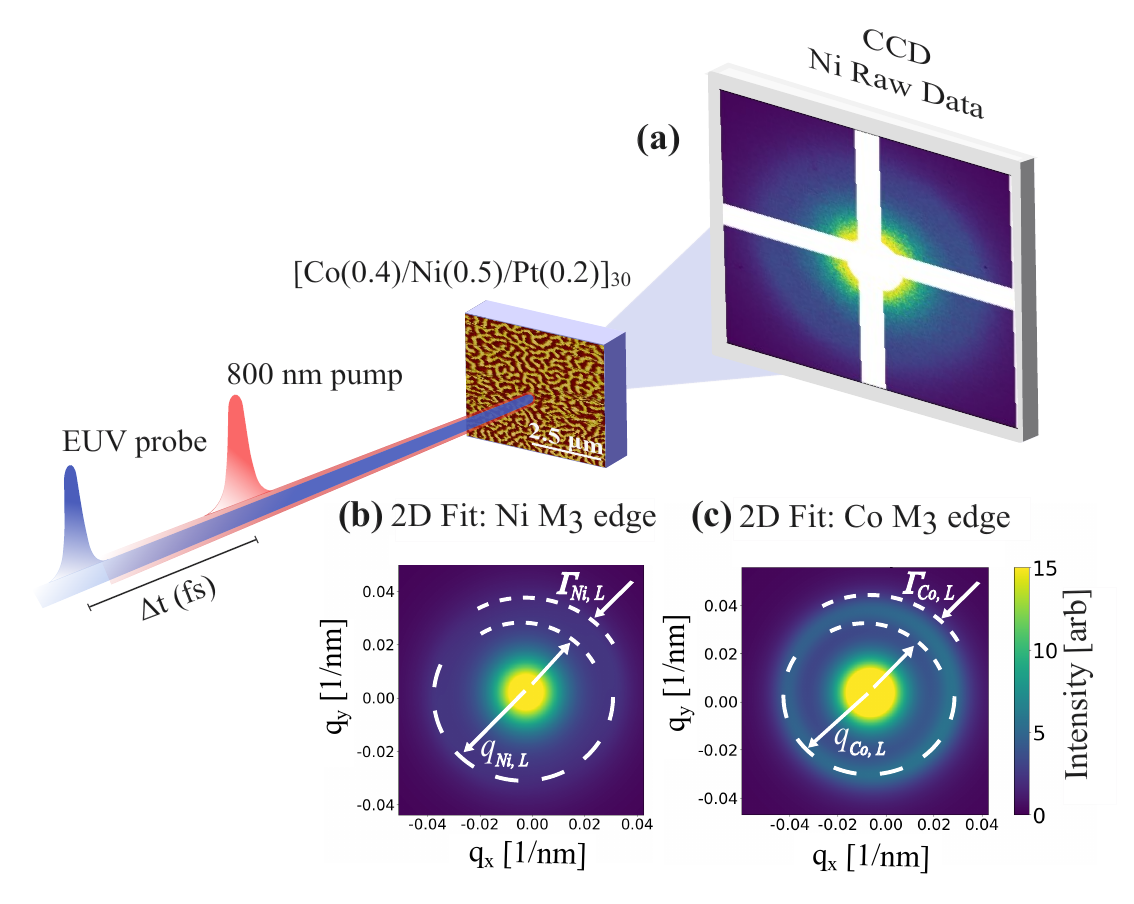}
    \caption{Experimental schematic for time-resolved magnetic scattering at DiProI beamline at FERMI: Optical pump and EUV magnetic scattering probe setup with the MFM image of the [Co/Ni/Pt] multilayer sample. (a) Experimentally measured 2D diffraction pattern at the Ni edge. Measured magnetic diffraction and low-q scattering were fit using a 2D phenomenological model as described in the text and shown in (b) for the Ni M-edge and (c) for the Co M-edge ($A_{Ni}$,$A_{Co}$), ($q_{Ni}$,$q_{Co}$), and ($\Gamma_{Ni}$,$\Gamma_{Co}$).}
    \label{fig:setup}
\end{figure}

Time-resolved small-angle magnetic resonant scattering experiments were performed on Ta(5nm)/Pt(3nm)/$\mathrm{[Co(0.4nm)/Ni(0.5 nm)/Pt(0.2 nm)]}_{30}$ at the FERMI free electron laser. The multilayer sample was deposited using DC magnetron sputtering on a single crystal (100) Si membrane (2 mm x 2 mm, 100 nm thick) etched from a Si substrate, allowing measurements in transmission geometry. The average domain width was 140 nm based on magnetic force microscopy (MFM) prior to the beamtime. Additional details of the magnetic properties of the multilayer samples can be found in Ref.~\cite{Brock}. Figure \ref{fig:setup} shows a schematic of the time-resolved magnetic resonant scattering experimental setup at the DiProI beamline at FERMI. The extreme ultraviolet (EUV) photon energy was tuned to the Co and Ni M$_{3}$ edges at 59.5 eV and 66.2 eV, respectively. An 800 nm ultrafast pulse laser was used for pumping, along with a pulsed linearly polarized EUV probe to perform time-resolved small-angle magnetic scattering. To ensure uniform dynamics in the probed region of the sample, the pump laser spot size was 390 $\upmu$m and the EUV spot size was 100 $\upmu$m. The laser pulse duration was 50 fs, and the measurements were performed at a repetition rate of 50 Hz. The temporal resolution was 60 fs. Due to experimental constraints, the scattering measurements at the Ni and Co edges were performed 1 mm apart on the same membrane.

The collected magnetic scattering pattern showed different features based on the nanoscale magnetic domain pattern of the illuminated region. The magnetic scattering pattern at the Ni edge exhibited an isotropic ring and azimuthally anisotropic lobes, characteristic of a mixed-state domain pattern with both labyrinthine and stripe components. Spontaneous formation of mixed domain states has previously been observed in similar samples due to plastic deformation of the supporting membrane due to heating by the pump laser \cite{ZhouHagstrom, Jangid}. No anisotropic scattering of lobes was observed at the Co-edge, likely due to different amounts of in-plane strain. Such spatial dependence of the diffraction symmetry was reported in \cite{ZhouHagstrom}. For quantitative analysis of the diffraction patterns, we utilized a phenomenological 2D fitting model similar to that described in \cite{ZhouHagstrom, Jangid}. The fitting process allows us to isolate the dynamics of the azimuthal-isotropic ring and the anisotropic lobes. Details of the fitting model are given in the Supplemental Material, Sec. A. Figure 1(b,c) presents the 2D fit results obtained at the Ni and Co edges at a fluence of 3.3 mJ/cm$^{2}$ for the isotropic ring and the anisotropic lobes that represent the labyrinth (L) and stripe components (S) of the domain pattern, respectively. For the Ni edge, the fitting parameters were amplitudes $A_{Ni,L}$ and $A_{Ni,S}$, radial peak positions $q_{Ni,L}$ and $q_{Ni,S}$, and widths $\Gamma_{Ni,L}$ and $\Gamma_{Ni,S}$. For the Co edge, the fitting parameters were those for the isotropic ring $A_{Co,L}$, $q_{Co,L}$, and $\Gamma_{Co,L}$ and an asymmetric component. Discussion of the asymmetric component for Co and low-q diffuse scattering for Co and Ni is included in the Supplemental Material, Sec. A, and is beyond the current scope. In this Letter, we discuss differences in the isotropic ring and anisotropic lobe dynamics between Co and Ni, representing scattering from the labyrinthine and stripe components of the domain pattern.

\begin{figure*}[t]
    \includegraphics[width=1.0\linewidth]{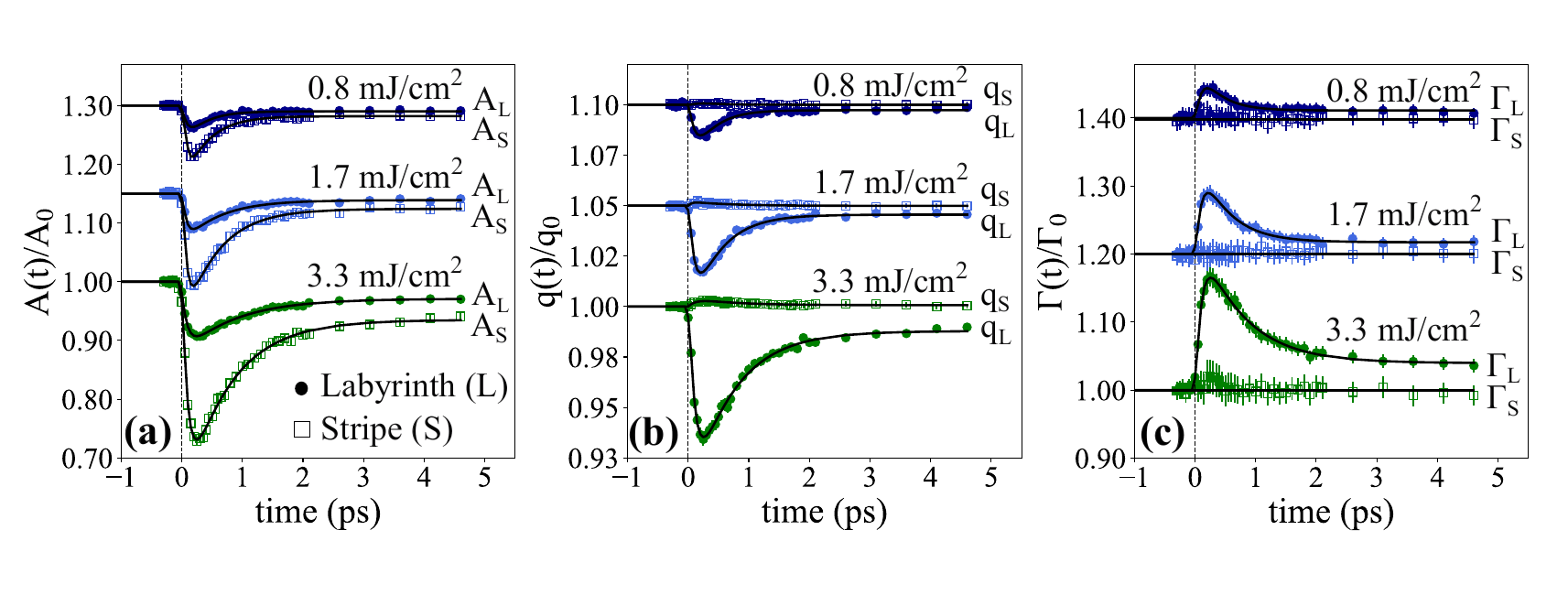}
    \caption{Ultrafast magnetization dynamics at the Ni M-edge: Delay curves for the isotropic ring and anisotropic lobe scattering originating from the labyrinth (L) and stripe (S) domains, including  (a) the amplitude ((A(t)/A$_0$)$_{Ni,L}$, (A(t)/A$_0$)$_{Ni,S}$), (b) radial position ((q(t)/q$_0$)$_{Ni,L}$,(q(t)/q$_0$)$_{Ni,S}$), and (c) radial width (($\Gamma$(t)/$\Gamma$$_0$)$_{Ni,L}$, ($\Gamma$(t)/$\Gamma$$_0$)$_{Ni,S}$) for a measured fluence range of 0.8 to 3.3 mJ/cm$^{2}$. The scattering amplitude, proportional to magnetization, undergoes quenching immediately following the laser excitation and shows recovery on ps timescales. The radial peak position and width approximate the average real-space domain size and correlation length of the respective domains. The exponential time-constant fits are overlayed as black curves. The vertical dashed line indicates t = 0. The delay curves for A, q, and $\Gamma$ are plotted relative to the value before t = 0.}
    \label{fig:NiDelay}
\end{figure*}

\begin{figure*}[t]
    \includegraphics[width=1.0\linewidth]{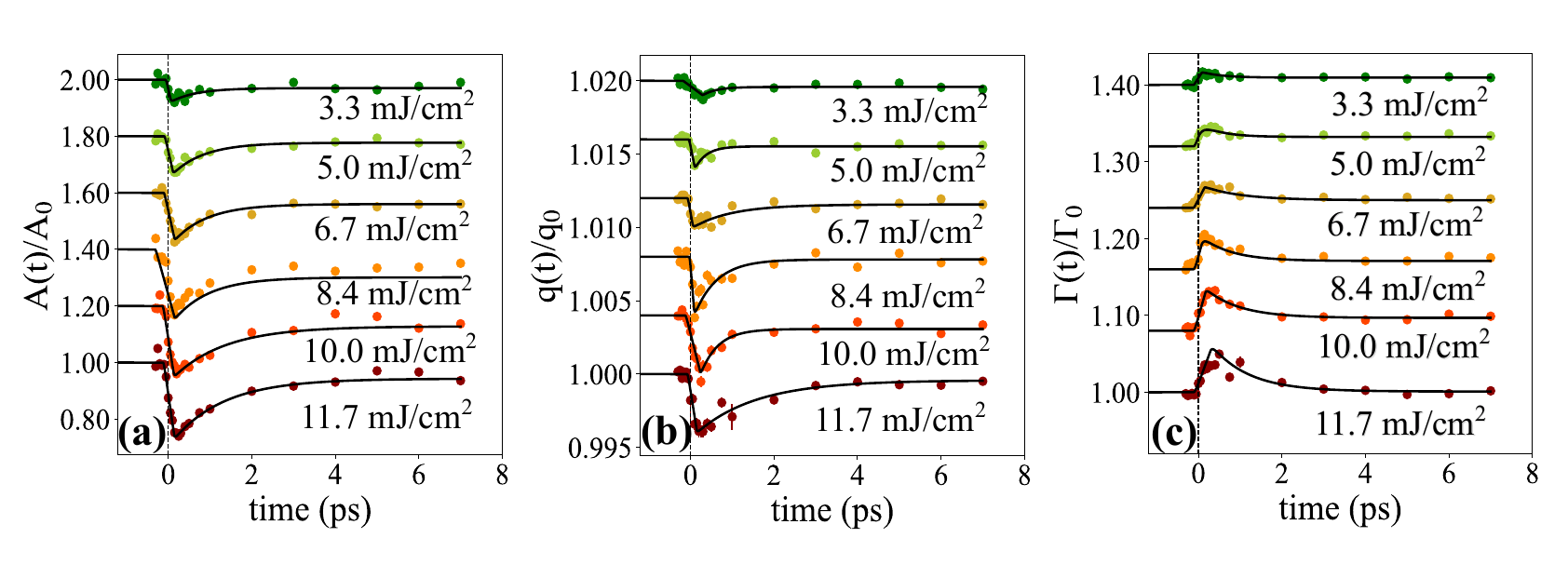}
    \caption{Ultrafast magnetization dynamics at the Co M-edge: Delay curves for the isotropic ring scattering due to labyrinth domains at the Co M-edge, including (a) the amplitude (A(t)/A$_0$)$_{Co,L}$, (b) radial peak position (q(t)/q$_0$)$_{Co,L}$, and (c) radial peak width ($\Gamma$(t)/$\Gamma$$_0$)$_{Co,L}$ for the measured fluence range of 3.3 to 11.7 mJ/cm$^{2}$. The exponential time-constant fits are overlaid as black curves. The vertical dashed line indicates t = 0. The delay curves for A, q, and $\Gamma$ are plotted relative to the value before t = 0.}
    \label{fig:CoDelay}
\end{figure*}

Figure \ref{fig:NiDelay} shows the temporal evolution for Ni magnetization amplitudes, radial peak positions, and widths for both the labyrinthine and stripe domain components of the mixed domain scattering pattern. Exponential time-constant fits are also included (black curves), and the fitting procedure is described in the Supplemental Material, Sec. B.  An ultrafast quench in the amplitude ($A_{Ni,L}$, $A_{Ni,S}$) is observed for both labyrinthine and stripe-like domains. The quench for the two domain types was found to be dissimilar. For the highest measured fluence at the Ni edge (3.3 mJ/cm$^{2}$), we obtained a magnetization quench of 9.28 $\pm$ 0.22\% for the labyrinthine domains and a quench of 26.91 $\pm$ 0.55\%  for stripe domains. The ratio of demagnetization between the labyrinthine and stripe domains slightly increased with fluence as 2.4 $\pm$ 0.3, 2.6 $\pm$ 0.1, and 2.9 $\pm$ 0.1, for 0.8 mJ/cm$^{2}$, 1.7 mJ/cm$^{2}$,  and 3.3 mJ/cm$^{2}$,  respectively. Another key difference between the labyrinthine and the stripe is the change in the diffraction features. For the fluence of 3.3 mJ/cm$^{2}$, a 6.47 $\pm$ 0.20\% and 16.52 $\pm$ 0.41\% modification of the labyrinthine ring position ($q_{Ni,L}$) and width ($\Gamma_{Ni,L}$) was observed, respectively, while the change in the stripe domain radial ring position ($q_{Ni,S}$) and width ($\Gamma_{Ni,S}$) was less than 0.5\%. The larger radial ring shift and broadening for the labyrinthine component, compared to the stripe component of the mixed domain pattern, is consistent with recent reports that attribute the curvature of the labyrinthine domains to ultrafast distortions following laser excitation \cite{ZhouHagstrom, Jangid}. This curvature in the labyrinthine domains provides the necessary symmetry breaking to drive ultrafast domain wall motion via spin transfer torque from superdiffusive spin currents \cite{Balaz, Jangid}. A recent ultrafast diffractive imaging study has also detected a domain wall displacement of 1.8 nm within 500 fs in TbCo at 6.2 mJ/cm${^2}$, but could not confirm wall motion due to measurement uncertainties \cite{Chang}. Similar imaging measurements are needed on ferromagnetic alloy systems to quantify the wall displacement and understand the dissimilar quench of stripe and labyrinth domains. 

\begin{figure*}[!t]
    \includegraphics[width=1.0\linewidth]{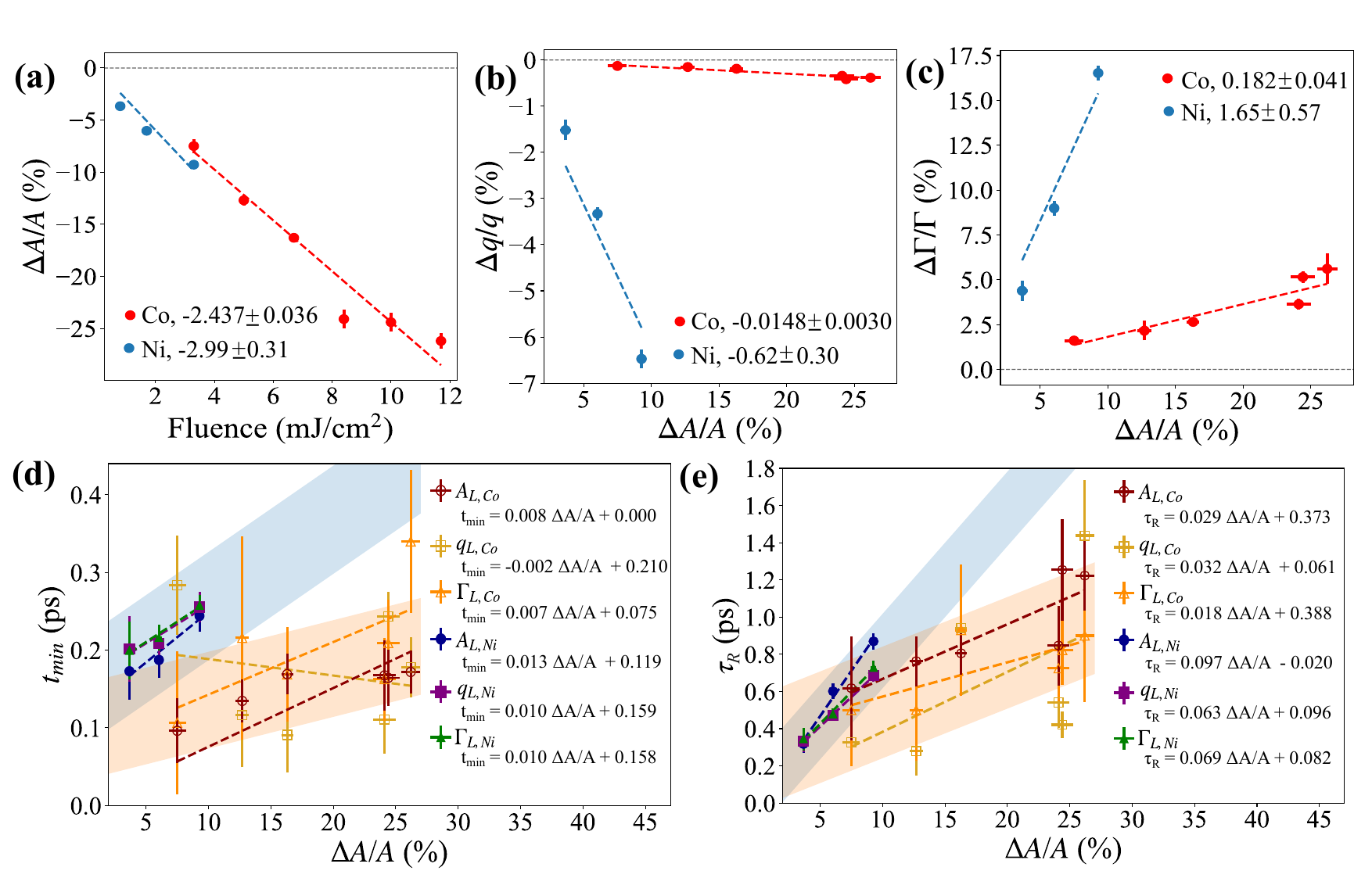}
    \caption{Comparison of magnetization dynamics in labyrinthine domains for Ni vs Co: (a) Fluence dependence of magnetization quench for Ni ($A_{Ni,L}$) and Co ($A_{Co,L}$). (b) Normalized radial shift ($\Delta$q/q)$_L$ and (c) width ($\Delta$$\Gamma$/$\Gamma$)$_L$ as a function of demagnetization amplitude ($\Delta$A/A), clearly showing that the modification of shift and width is smaller for Co compared to Ni. For plots (a) to (c), a linear regression model was used with constrained intercepts (i.e., q=$\Gamma$=0 at $\Delta A/A$=0) and the value of slope, including its error, is reported in the legend. (d) Quench and (e) recovery time constants for both Ni and Co obtained from the temporal fits for $A_{Ni,L}$, $A_{Co,L}$, $q_{Ni,L}$, $q_{Co,L}$, $\Gamma_{Ni,L}$ and $\Gamma_{Co,L}$. The dashed line in all plots indicates the slope fit with a linear regression model. The slopes and intercepts are reported in the legend of each plot. More details on fitting can be found in Supplemental Material, Sec. C.}
    \label{fig:fluence}
\end{figure*}

Figure \ref{fig:CoDelay} shows the temporal evolution of the labyrinthine-type domains at the Co-edge including $A_{Co,L}$, $q_{Co,L}$, and $\Gamma_{Co,L}$, for the fluence range of 3.3 to 11.7 mJ/cm$^{2}$. The vertical scale of delay curves for $q_{Co,L}$ and $\Gamma_{Co,L}$ presented in Figure \ref{fig:CoDelay}(b) and (c) indicates that the magnitude of the ultrafast modification of the radial ring shift and width is smaller in Co than Ni. To directly compare the magnitude of labyrinthine domain magnetization quench, radial ring shift and width for Co and Ni, Figure \ref{fig:fluence}(a)-(c) presents the maximum ultrafast modification normalized to its value before t=0, i.e. $\Delta A/A$, $\Delta q/q$, and $\Delta \Gamma / \Gamma$. Slopes obtained from linear regression fits for each parameter are also included in the legend. 

Figure \ref{fig:fluence}(a)-(c) clearly illustrates that the modification of $\Delta$q/q and $\Delta$$\Gamma$/$\Gamma$ are significantly lower for Co than for Ni. In fact, $\Delta$q/q is an order of magnitude smaller for Co compared to Ni for the same magnetization quench ($\Delta A/A$). Furthermore, the slope of $\Delta$q/q for Ni is 41 times larger than Co, and the slope of $\Delta$$\Gamma$/$\Gamma$ for Ni is 9 times larger than Co. Note that in Figure \ref{fig:fluence}(b) and (c), $\Delta A/A$ was used to directly compare the ultrafast distortions in ring radius and width, instead of fluence, as magnetization quench for Ni and Co are similar for a given fluence. The slightly larger slope of $\Delta A/A$ for Ni compared to Co (1.2 times) is consistent with a previous study where simulations of Co/Pt and Ni/Pt interfaces indicated that Ni/Pt interfaces quench slightly more ($\sim$1.13) than Co/Pt interfaces \cite{Dewhurst}. This slight difference in magnetization quench was attributed to variations in the density of states and spin-orbit coupling of the Pt interfaces \cite{Dewhurst}. Nevertheless, our results clearly reveal that in spite of similar magnetization quench for Co and Ni, the element-specific response of $\Delta$q/q and $\Delta$$\Gamma$/$\Gamma$, indicative of distortions in the domain pattern, is distinct between Ni and Co. In other words, the demagnetization dynamics of magnetic textures for Ni and Co subsystems within a thin film are notably different. 

To further investigate the distinct magnetization dynamics of Ni and Co, we compared the demagnetization timescales, including quench and recovery time-constants obtained at the Ni and Co edges. Figure 4(d)-(e) plots the quench ($t_{min}$) and recovery ($\tau_R$) time-constants for $A_{Ni,L}$, $A_{Co,L}$, $q_{Ni,L}$,  $q_{Co,L}$, $\Gamma_{Ni,L}$ and $\Gamma_{Co,L}$. Details on the time constant fitting are provided in the Supplemental Material, Sec B. Note that $\Delta A/A$ (instead of fluence) was used for Figure \ref{fig:fluence}(d) and (e) to directly compare the behavior of Ni and Co. The dashed lines represent the linear regression fits with parameters included in the legend. The time constant fits for Ni were found to be similar for $A_{Ni,L}$,  $q_{Ni,L}$ and $\Gamma_{Ni,L}$ with an average $t_{min}$ of 0.252 $\pm$ 0.004 ps and $\tau_R$ of 0.760$\pm$0.050 ps at an incident fluence of 3.3 mJ/cm$^{2}$. The average $\tau_R$ for all three fluences is 536 $\pm$ 65.5 fs, or approximately 500 fs.  The reported $t_{min}$ and $\tau_R$ are similar in magnitude to those reported in \cite{ZhouHagstrom,Jangid}. Larger variation in time constants was observed for $A_{Co,L}$, $q_{Co,L}$, and $\Gamma_{Co,L}$ at the Co edge. The higher error in Co fits is due to a significantly lower magnitude of modification for Co $q_{Co,L}$ and $\Gamma_{Co,L}$ compared to Ni. For Co, the average $t_{min}$ was 0.162$\pm$0.048 ps and $\tau_R$ was 0.481$\pm$0.069 ps at the incident fluence of 3.3 mJ/cm$^{2}$. Overall, the average slope for Ni fit parameters was found to be 2.5 and 2.9 times greater than the Co edge $t_{min}$ and $\tau_R$, respectively. Our results highlight that Ni shows slightly slower quench and recovery, further indicating the distinct response of Co and Ni in the thin film.

In order to explain the surprising difference between Co and Ni layers, we considered laser-driven modification of the exchange energy for Ni/Pt and Co/Pt interfaces, which can manifest as a variation in the angle of neighboring spins in the adjacent layers (i.e. Ni and Co spins). Recent studies have proposed that modified exchange interactions may explain element-specific magnetization dynamics in magnetic alloys \cite{Graves,Vaskivskyi}. Furthermore,  it was recently shown that ultrafast variation of the exchange stiffness can strongly influence recovery time scales in [Co/Pt] multilayers \cite{Shim2020}. The experimentally obtained average recovery time of $\sim$500 fs for Ni implies that the recovery process has a characteristic energy of 4.14 meV. The interfacial exchange energy of [Co/Ni/Pt] is around 0.09 eV \cite{Talagala}. We utilized these two energies to calculate a canting angle of 18\degree between neighboring spins across the Co and Ni interface (details provided in the Supplemental Material, Sec. C). These values are also comparable to recent reports on the ultrafast precession angle ($\sim$10 degrees) of chiral domain walls in Co/Pt multilayers \cite{Léveillé}, considering different experimental factors such as incident fluence, sample, geometry, etc. Thus, the variations in canting angle between Co and Ni are consistent with the fast recovery times observed here. The modified interfacial exchange interactions, due to laser excitation, can drive the realignment of Co and Ni layers, leading to the observed fast recovery time scales.

Ultrafast exchange-driven distortions can also give rise to magnon generation, influencing the observed element-specific behavior and the ultrafast time scales of magnetization damping and recovery. In our films, the magnon wavelength is dictated by the periodicity of each layer in the multilayer (11 \AA \ or 0.6 \AA$^{-1}$). Wavenumbers greater than  0.1 \AA $^{-1}$ imply strongly damped magnons with energy overlapping Stoner excitations (spin flips of the conduction band electrons). Inelastic neutron scattering studies have shown that the magnon linewidth broadening in this limit (q=0.19 \AA$^{-1}$) is 20 meV for Ni and has a lifetime of 207 fs  (see Supplemental Material, Sec. D for additional calculations)  \cite{Cooke}. Thus, the average measured recovery time of $\sim$500 fs in our films is on the order of the estimated magnon lifetime in Co and Ni, indicating magnon fluctuation and dissipation\cite{Mook,Cooke,Zhang}. In fact, laser-driven wall fluctuations have been reported in [Co/Pt] \cite{Suturin}, and oscillations of domain walls leading to magnon generation have been observed by Lorentz transmission electron microscopy (L-TEM) in a Ni$_{80}$Fe$_{20}$ alloy \cite{Liu2025}.

Furthermore, the inclusion of Pt could also strongly influence the ultrafast modification of the exchange interactions or magnon generation. For example, a comparison with a [CoFe/Ni] multilayer measured by Jangid et al. \cite{Jangid}, indicates that the [Co/Ni/Pt] studied here has a lower magnetization quench and larger domain pattern distortions. Specifically, we measured a maximum q-shift of 6.5\% in Ni for a quench of 9.3\% (3.3 mJ/cm$^2$), whereas the maximum q-shift reported in Jangid et al. was 5.5\% for a quench of 40\% (13 mJ/cm$^2$) \cite{Jangid}. In other words, we find a comparable q-shift for Ni in [Co/Ni/Pt] and [CoFe/Ni] multilayers, but for a magnetization quench that is almost four times smaller. Additionally, ferromagnetic interfaces with heavy non-magnets (Pt) induce orthogonal  interfacial Dzyaloshinskii–Moriya interactions (iDMI), leading to chiral domain wall textures. Kerber et al. \cite{Kerber2020} and L\'eveill\'e et al. \cite{Léveillé} found that the chiral order of the domain walls recovers faster than the ferromagnetically aligned domains, resulting in distortions of the domain walls at ultrafast timescales. In [Co/Ni/Pt], the distortion of the chiral order in the vicinity of Co/Pt and Ni/Pt interfaces could also lead to the observed differences in Co and Ni. Our measurements demonstrate that the inclusion of Pt greatly enhances the efficiency of the spatial spin texture distortion driven by pumped magnons and laser excitation.

In conclusion, our measurements reveal distinct dynamics in Co and Ni, suggesting a complex 3D modification of the domain pattern along the film thickness. The recovery timescales for the diffraction pattern distortions in Ni and Co are fast, within 300-800 fs. The fast recovery timescales are consistent with modification of exchange interactions \cite{Shim2020,Vaskivskyi,Graves}, or magnon excitation and relaxation \cite{Zhang,ZAKERI2014,Elhanoty,TurgutPRB,Beens,Léveillé, Knut, Choi2014}. Our studies clearly illustrate that measuring the element-specific response is essential for a comprehensive understanding of spin and energy transfer across interfaces in the far-from-equilibrium regime. Our study also emphasizes that the incorporation of different layers and their distinct spin-orbit scattering strengths could offer a new approach for manipulating the magnitude and timescales of domain pattern distortions, which would be required for laser-driven spintronic applications.

The work performed was supported by AFOSR Grant. No. FA9550-23-1-0395. E. I. acknowledges support from the National Science Foundation under Grant No. 2205796. S. B. acknowledges support from the Italian Ministry of University and Research, PRIN2020 funding program, Grant No. 2020PY8KTC. The authors acknowledge the FERMI Free Electron laser in Trieste, Italy, for allowing us to use the Diffraction and Projection Imaging (DiProI) beamline and thank the beamline scientists and facility staff for their assistance. This research was funded by the National Institute of Standards and Technology (NIST). The identification of specific equipment, instruments, software, or materials in this paper does not imply endorsement by NIST, nor does it indicate their superiority for the intended purpose.

\bibliography{CovNi}

\supplementarysection

\subsection{Phenomenological 2D Fits and Results}
\begin{figure}[hbt!]
    \includegraphics[width=1.0\linewidth]{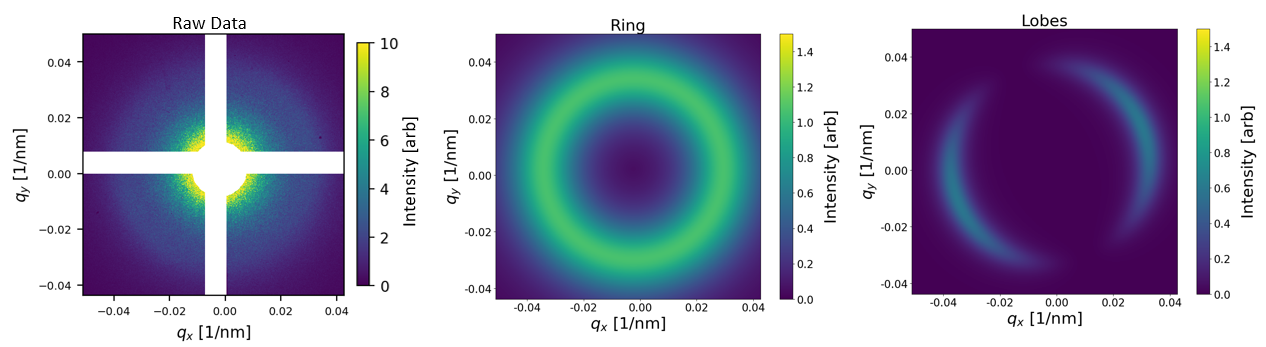}
    \caption{Phenomenological 2D fits for Ni including the raw data, isotropic ring, and anisotropic lobe scattering components. The full 2D fit is presented in Figure 1 of the main text.}
    \label{fig: NiHarmonics}
\end{figure}

\begin{figure}[hbt!]
    \includegraphics[width=1.0\linewidth]{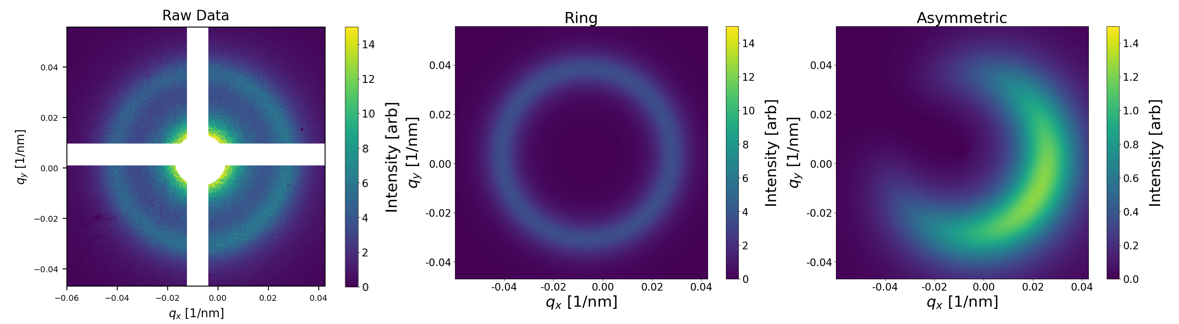}
    \caption{Phenomenological 2D fits for Co including the raw data, isotropic ring, and anisotropic lobe scattering components. The full 2D fit is presented in Figure 1 of the main text.}
    \label{fig:CoHarmonics}
\end{figure}

Phenomenological 2D fit equations were used to fit the scattering pattern obtained by the charge-coupled device (CCD) at the Co and Ni edges. The 2D fitting procedure, developed by \cite{ZhouHagstrom, Jangid} for the Fourier components (0$^{th}$, 1$^{st}$, and 2$^{nd}$ order) of scattering, was expanded to include low-q diffuse scattering. The fits were modeled using symmetric Lorentzians for the diffuse (low-q), isotropic ring (0$^{th}$ order), asymmetric (1$^{st}$ order), and anisotropic lobe (2$^{nd}$ order) scattering components. For the Ni edge, the fit included diffuse (D), isotropic ring (L), and anisotropic lobes (S) scattering components given by, 
\begin{equation}\label{Ni}
\begin{split}
 f_{Ni}(I, q,\Gamma, \phi ) = B + f_{D}(I_{D, Ni}, \Gamma_{D, Ni} ) \\ + f_{L}(I_{L, Ni}, q_{L, Ni}, \Gamma_{L, Ni} ) \\ + f_{S}(I_{S, Ni}, q_{S}, \Gamma_{S, Ni}, \phi_{S, Ni} )
\end{split}
\end{equation}

where q is the wave vector peak position, $\phi$ is the azimuthal angle, $\Gamma$ is the linewidth, B is the background constrained to the pre-pump average, and I is the intensity of its respective component. For the Co edge, the fit included diffuse (D), isotropic (L), and asymmetric (as) scattering. No anisotropic scattering was observed from stripe-like domain components, potentially because of the different measurement locations on the same membrane for the Co vs Ni edge. The discussion of the diffuse peak is beyond the scope of this Letter. The fit equation at the Co edge is given by, 
\begin{equation}\label{Co}
\begin{split}
 f_{Co}(I, q,\Gamma, \phi ) =  B + f_{D}(I_{D, Co}, \Gamma_{D, Co} ) \\ + f_{L}(I_{L, Co}, q_{L, Co}, \Gamma_{R, Co}) \\ + f_{as}(I_{as, Co}, q_{as, Co},\Gamma_{as, Co} , \phi_{as, Co})
 \end{split}
\end{equation}
The full equation for each fit function in Equation \ref{Ni} and \ref{Co} is as follows. Note that the same equations were used for both elemental edges. 

\begin{equation}
f_{D}(I_{D}, \Gamma_{D} ) = |I_{D}| {\left[  \frac{ 1 } { \left( \frac{q^2}{\Gamma_{D}^2} \right)  + 1 }  \right]}^2
\end{equation}
   \begin{equation}
f_{L}(I_L, q_L, \Gamma_L ) = |I_{L}| {\left[  \frac{ 1 } { \left( \frac{(q-q_L)^2}{\Gamma_{L}^2} \right)  + 1 }  \right]}^2
\end{equation}

\begin{equation}
\begin{split}
 f_{as}(I_1, q_{as},\Gamma_{as} , \phi_{as} ) = {\left[  \frac{ 1 } { \left( \frac{(q-q_{as})^2}{\Gamma_{as}^2} \right)  + 1 }  \right]}^2 \\* \left(  \frac{|I_1|}{2} * (cos(\theta - \phi_{as} ) +1) \right)
\end {split}
\end{equation}

\begin{equation}
\begin{split}
 f_{S}(I_S, q_{S},\phi_{S} ) = {\left[  \frac{ 1 } { \left( \frac{(q-q_{S})^2}{\Gamma_{S}^2} \right)  + 1 }  \right]}^2 \\ * \left(  \frac{|I_S|}{2} * (cos(2\theta - 2\phi_{S} ) +1) \right)
 \end{split}
\end{equation}
The general fitting procedure to determine the necessary fitting components for Co and Ni included (i) determining the center of scattering. (ii) Fitting the diffuse peak from 1D azimuthal integration. (iii) Fitting the isotropic ring intensity from the 2D azimuthal integration. (iv) Assessing the residual for harmonic components, either asymmetric ($1^{st}$) or lobe ($2^{nd}$). (v) Inputting the initial guesses determined in steps (i)-(iv) to the full phenomenological fit to output the best fit for the eight parameters at each time delay.  Figure \ref{fig:setup} in the main text shows the full 2D fit for Co and Ni at 3.3 mJ/cm${^2}$ obtained using the discussed fitting process. The individual fits of the respective $0^{th}$, $1^{st}$, and $2^{nd}$ harmonics for Co and Ni are shown in Figures \ref{fig: NiHarmonics} and \ref{fig:CoHarmonics}.

\begin{figure}[hbt!]
    \includegraphics[width=1.0\linewidth]{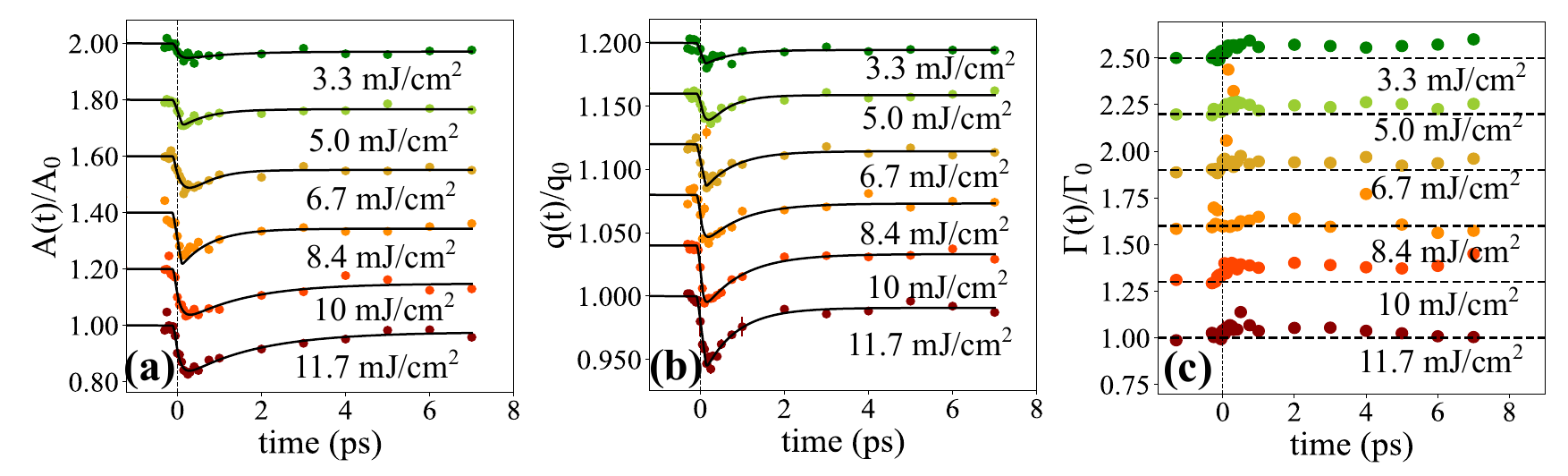}
    \caption{Evolution of the odd harmonic scattering observed at the Co edge. (a) scattering amplitude ($A_{as}$), (b) radial peak position ($q_{as}$), and (c) radial peak width ($\Gamma_{as}$) plotted as a function of delay time for all measured pump fluences. Black curve represents time constant fits, which were obtained as described in the text.}
    \label{fig:CoAsymmetric}
\end{figure}

\subsection{Time Constant Fits and Results}
The following fitting procedure was used to obtain time constants for the ultrafast behavior of each parameter in the phenomenological 2D fit (amplitude, radial peak position, and width). The fit equation is, 
\begin{equation}\label{Forcing}
\begin{split}
M_{0}(t, \tau_{m}, \tau_{R}, M_{q}, M_{R}) = \\ \mathrm{if } \:  (t_{0} < t < |\tau_{m}|) \left [1 - \left(  \frac{ 1-M_q } {|\tau_{m}|} * t  \right ) \right],  \\ \mathrm{else } \: (t> |\tau_{m}|)\left [ M_{R} + (M_{q}-M_{R})*exp\left \{ - \frac{t - |\tau_{m}|}{|\tau_{R}|}\right \} \right] 
\end{split}
\end{equation}
where $\tau_{m}$ is the quench time constant, $M_{q}$ is the effective quench value, $\tau_{R}$ is the recovery time constant, and $M_{R}$ is the recovered magnetization. Below time zero ($t_{0}$ = 0 ps) we assume $M_{0}$ is equal to 1. Due to the fast quench, $M_{0}$ is considered to quench to $M_{q}$ between times $t_{0}$ and $\tau_{m}$ as passing a forcing function through a low-pass filter. Recovery was fit using an exponential function.

Equation \ref{Forcing} was solved analytically with a first-order ordinary differential equation (ODE) as a piece-wise forcing function. For the linear quench to $M_q$, the initial guess of the particular solution was a first-order polynomial, resulting in the general solution given by, 
\begin{equation}
\begin{split}
 M_{1}(t) = \\ 1 - \left( \frac{1-M_{q}}{|\tau_{m}|} \right)t+\left( \frac{1-M_{q}}{|\tau_{m}|} \right)* \\\tau*\left(1-exp\left \{-\frac{t}{|\tau|}\right \} \right) 
 \end{split}
\end{equation}

For the exponential recovery to $M_R$, the general solution, $M_2$,  included an exponential and a constant term B, which captures the long-term recovery and enforces continuity at $M_1$ and $M_2$ with the smoothing term $\tau$,

\begin{equation}
\begin{split}
 M_{2}(t) =  B*exp\left \{-\frac{t  - \tau_{m}}{\tau}\right \} + M_{R} +\\  (M_{q} - M_{R})* \left( \frac{\tau_{R}}{|\tau_{R}-\tau|} \right)*exp \left \{-\frac{t -\tau_{m}}{\tau_{R}}\right \} 
 \end{split}
 \end{equation}
 \begin{equation}
 \begin{split}
 B = M_{q} - M_{R} - (M_{q} - M_{R})*\left( \frac{\tau_{R}}{|\tau_{R}-\tau|} \right)  \\ + \left( \frac{1-M_{q}}{|\tau_{m}|} \right)*\tau* \left ( 1 - exp\left \{-\frac{\tau_{m}}{\tau}\right \} \right)
 \end{split}
\end{equation}
The minimum of the fit occurs in $M_2$ where a derivative of $M_2$ equal to zero will yield the local minima related to $M_{min}$ and $t_{min}$. This fitting procedure was used for both the Ni and Co resonant edges. Table \ref{tab:TCF Table} shows all time constant fit slopes, intercepts, and the Pearson regression coefficient R for the linear regressions plotted in Figure \ref{fig:fluence}(d) and (e).

\begin{table*}[!t]
    \caption{Slope, intercept, and regression coefficient for quench time (t$_{min}$) and recovery time ($\tau_R$) obtained in Section C. All errors were calculated using standard error propagation.}
    \includegraphics[width=1.0\linewidth]{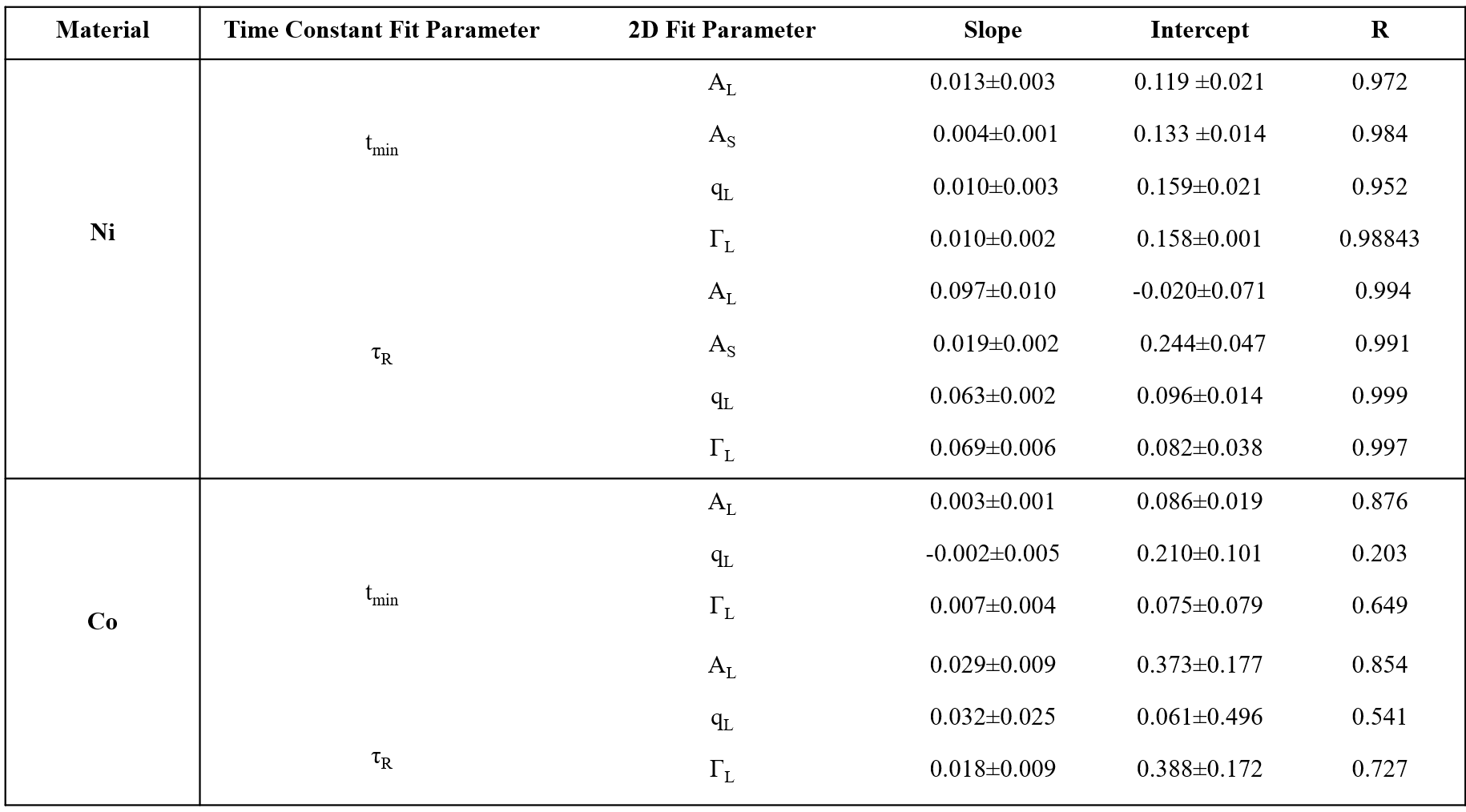}
    \label{tab:TCF Table}
\end{table*}

\subsection{Calculations based on the modification of exchange energy}
\begin{figure}[hbt!]
    \includegraphics[width=1.0\linewidth]{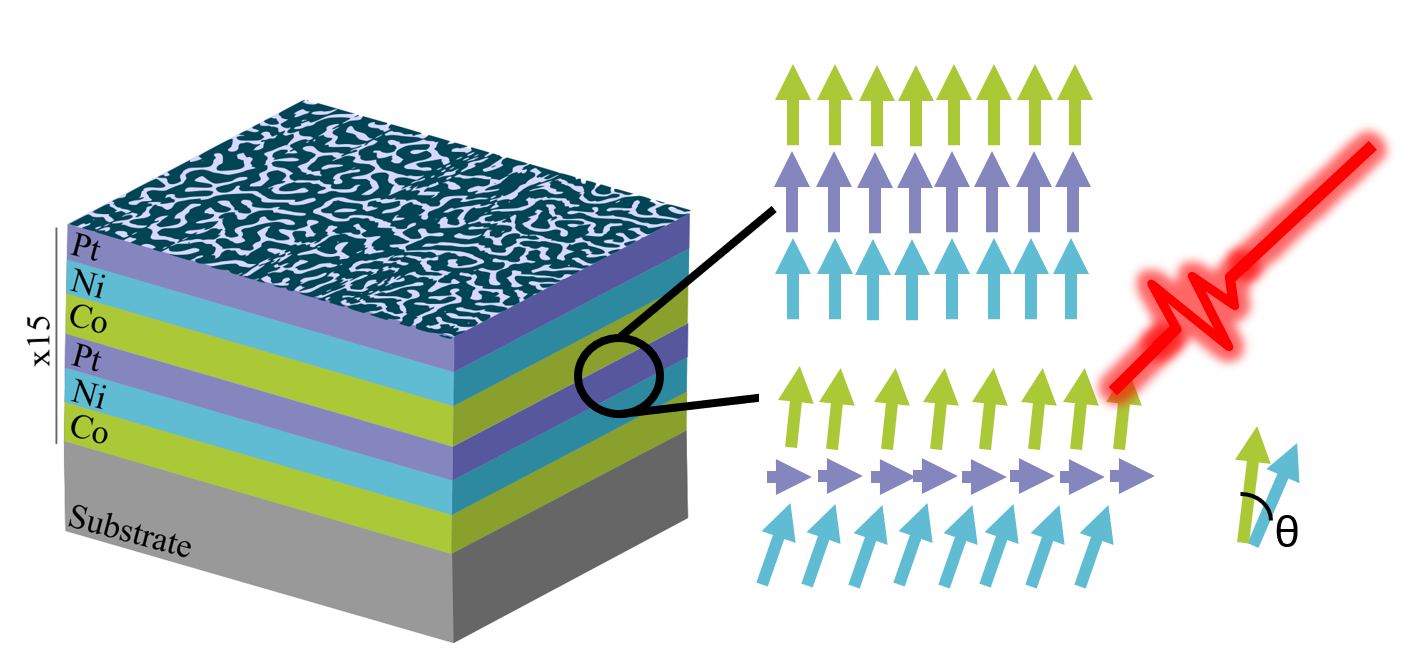}
    \caption{Schematic of the [Co/Ni/Pt] multilayer with the MFM image overlayed on top. Assumed spin alignment before and after pumping (red) is shown, where the calculated canting angle of 17.56\degree was calculated between Co (green) and Ni (blue).}
    \label{fig:multilayerCanting}
\end{figure}

Modifications of the exchange energy (E$_{ex}$) at the Ni/Co interface via the angle, $\theta$, between two adjacent spins after laser excitation are considered for the average recovery time of 500 fs. A schematic is shown in Figure \ref{fig:multilayerCanting}. The characteristic energy (E$_{char}$) associated with such a fast recovery time ($\tau$) is 4.13 meV i.e. frequency (f) of $\sim$1 THz ($f=\frac{1}{2\tau}$). The relative thickness of the interface compared to the bulk ($\epsilon$) is approximated as, 

\begin{equation}
\epsilon = \frac{a } {d_{Ni} + d_{Co} }
\end{equation}
where a is the lattice constant of 0.35 nm \cite{Liu}, Ni thickness ($d_{Ni}$) and Co thickness ($d_{Co}$) are 0.5 nm and 0.4 nm, respectively. This ratio was used to estimate the interfacial exchange ($D_{int}$) averaged over the layer thickness to get the effective exchange stiffness,
\begin{equation}
D = \epsilon * D_{int}  
\end{equation}
$D_{int}$ is equal to 0.5 eV \r{A}$^2$ from \cite{Talagala}. Using this value for spin wave dispersion at the zone boundary, 
\begin{equation}
E_{int} =  D * k^2 , k =\frac{ \pi } {a } 
\end{equation}
The exchange energy associated with a change in magnetization direction at the interface was 0.088 eV. This value is similar to the $\sim$0.1 eV exchange energy reported in \cite{Leveille2022}, where hot electrons excited by the 1 eV excitation were expected to have lower exchange energy compared to electrons near the Fermi level. Exchange energy is proportional to the difference in angle ($\theta$) between the two sides of the interface, 
\begin{equation}
\frac {E_{char}}{E_{int}}= 1- cos(\theta)
\end{equation}
and when considering a Taylor series expansion yields,
\begin{equation}
    \frac{\theta^2}{2} = \frac {E_{char}}{E_{int}}
\end{equation}
resulting in,
\begin{equation}
    \theta = \sqrt{\frac {2E_{char}}{E_{int}}}
\end{equation}
Solving for $\theta$, a canting angle of 17.56\degree  is obtained for the Ni/Co interface at 500 fs. This canting angle represents the angle between neighboring spins across the interface needed to achieve the observed recovery time as discussed in the main text. Thus, considering a laser-driven modified interfacial exchange interaction is consistent with the fast recovery $\tau_R$.  

\subsection{Calculations based on Magnon Excitation}
As discussed in the main text, the variation of the magnetization through the multilayer thickness at q>0 could be due to a magnon with the wave vector along the interface normal. The magnon wavelength is set by the thickness of each triplet in the multilayer (1.1 nm or 0.6 \AA$^{-1}$).  This wavelength is short. Thus, the damping could be due to Landau damping, where the magnons are strongly damped when the energy overlaps with Stoner excitations (i.e, spin flip excitations of the spin split conduction bands) at wavenumbers greater than 0.1 \AA$^{-1}$. To explain the differences in the radial peak shift and broadening between Co and Ni, we consider exchange energy (E) (dependent on the exchange constant, $J_{ex}$, and the angle between two adjacent spins, $\theta$, in Supplemental Material, Sec. C) and then considered the magnitude of the torque (T) acting on each of the spins due to E. 
\begin{equation}
E = \frac{J_{ex}} {2} \theta ^ {2}  
\end{equation}
\begin{equation}
T = J_{ex} \theta 
\end{equation}
In the high-damping limit, the damping torque (T$_d$) and exchange torque (T) are approximately antiparallel,
\begin{equation}
T_d = -\nu(T_m) \dot{\theta} = -\tau_{r,m}(T_m) J_\text{ex} \dot{\theta}
\end{equation}
Damping is proportional to the magnon temperature, $T_{m}$, and $\tau_{r,m}$ is the magnon lifetime. The equation of motion for two neighboring spins can be approximated from $T_d$ as, 
\begin{equation}
\theta = \theta_0 e^{-\frac{t}{\tau_{r,m}(T_m)}}
\end{equation}
For a magnon linewidth broadening at q=0.19  \AA$^{-1}$ with 20 meV in Ni, the estimated magnon lifetime is 207 fs \cite{Cooke}.  Furthermore, both inelastic neutron scattering \cite{Mook} and spin-polarized electron energy loss spectroscopy (SPEELS) indicate that the magnon lifetimes are on the order of 100 fs \cite{Zhang}. This agrees with our fastest measured recovery time ($\tau_R$) of 300 fs for Co and Ni, where the dependence of $\tau_R$  may be a manifestation of magnon fluctuation and dissipation. Thus, the modification in the Ni spatial pattern relative to Co could be due to magnon propagation, which sets the ultrafast time scales for magnetization damping and recovery.

\end{document}